\documentclass[nofootinbib,onecolumn]{revtex4}
\pdfoutput=1 
\usepackage{graphicx}
\usepackage{rotate}
\usepackage{amsmath}
\usepackage{amssymb}
\usepackage{amsfonts}
\usepackage{enumerate}
\usepackage{afterpage}
\usepackage{xcolor}
\usepackage{natbib}
\usepackage{bm}
\usepackage{hyperref}

\renewcommand{\mathbf}{\bm}
\renewcommand{\k}{\bm k}
\newcommand{\ud}{\mathrm{d}}

\renewcommand{\H}{\mathcal{H}}
\newcommand{\fnl}{f_\text{NL}}

\def\be{\begin{equation}}
\def\ee{\end{equation}}
\def\bea{\begin{eqnarray}}
\def\eea{\end{eqnarray}}

\begin{document}

\title{A general relativistic signature in the galaxy bispectrum:\\ {the local effects of observing on the lightcone}}

\author{Obinna Umeh$^1$, Sheean Jolicoeur$^1$, Roy Maartens$^{1,2}$, Chris Clarkson$^{3,4}$}

\affiliation{~\\ $^1$Department of Physics \& Astronomy, University of the Western Cape, Cape Town 7535, South Africa\\
$^2$Institute of Cosmology \& Gravitation, University of Portsmouth, Portsmouth PO1 3FX, UK\\
$^3$School of Physics \& Astronomy, Queen Mary University of London, London E1 4NS, UK\\
$^4$Department of Mathematics \& Applied Mathematics, University of Cape Town, Cape Town 7701, South Africa
}
\date{\today}

\begin{abstract}

Next-generation galaxy surveys will increasingly rely on the galaxy bispectrum to improve cosmological constraints, especially on primordial non-Gaussianity. A key theoretical requirement that remains to be developed is the analysis of general relativistic effects on the bispectrum,  which arise from observing galaxies on the past lightcone, {as well as from relativistic corrections to the dynamics}. {As an initial step towards a fully relativistic analysis of the galaxy bispectrum,  we compute for the first time the local relativistic lightcone effects on the bispectrum,}  which come from Doppler and gravitational potential contributions. For the galaxy bispectrum, the problem is much more complex than for the power spectrum, since we need the lightcone corrections at second order.  Mode-coupling contributions at second order mean that relativistic corrections can be non-negligible at smaller scales than in the case of the power spectrum. 
In a primordial Gaussian universe, we show that the local lightcone corrections for squeezed shapes at $z\sim1$ mean that the bispectrum can differ from the Newtonian prediction by $\gtrsim 10\%$ when the short modes are $k\lesssim (50\,{\rm Mpc})^{-1}$.
These relativistic projection effects, if ignored in the analysis of observations, could  be mistaken for primordial non-Gaussianity. 
For upcoming surveys which probe equality scales and beyond, {all relativistic lightcone effects and relativistic dynamical corrections should be included} for an accurate measurement of primordial non-Gaussianity. 

\end{abstract}
\maketitle

\subsection*{ Introduction}

With the coming generation of large-scale surveys, the galaxy bispectrum will play an increasingly important role, supplementing the power spectrum. In particular, the bispectrum will be crucial  in discriminating inflationary models via constraints on primordial non-Gaussianity \cite{Tellarini:2016sgp}.

It is well known that the primordial signal is contaminated by nonlinearity in the matter overdensity and in the galaxy bias, {which requires accurate modelling for cosmological constraints and forecasts.} Here we focus on a lesser known source of non-Gaussianity, i.e. the `projection' effects that arise from observing on the past lightcone. These general relativistic effects are inescapably present in the measurements and lead to a contamination of the primordial signal. The standard Newtonian analysis of the bispectrum incorporates the projection effect due to redshift-space distortions, but omits all other projection effects, {as well as the relativistic corrections to nonlinear dynamics.} {For next-generation surveys, the Newtonian approximation cannot be expected to deliver the necessary theoretical accuracy.}

Recently, \cite{DiDio:2015bua} computed the effects of lensing on the angular bispectrum, showing that this relativistic correction to the Newtonian analysis is relevant at higher redshifts and on intermediate scales. 
{They neglected all other relativistic effects: local projection effects,  further integrated projection effects (from integrated Sachs-Wolfe and time delay effects and their couplings), and relativistic dynamical corrections.
Our work is complementary to theirs, in the sense that we include all local relativistic projection effects, while neglecting the remaining effects.
This allows us to use the standard Fourier space approach to the bispectrum (see e.g. \cite{Tellarini:2016sgp,Verde:1998zr,Baldauf:2010vn,Gil-Marin:2016wya}). A consequence of the standard Fourier space analysis is that the plane-parallel approximation is adopted, which means that the contribution of wide-angle correlations is missed. Some of our results on ultra-large scales will be affected by this.}  
 
{Even without nonlocal projection effects and dynamical corrections, the computation is complicated,} and the details of the derivation are given in an accompanying paper \cite{Jolicoeur:2016}. We use the  expression up to second order for relativistic effects on the observed galaxy number counts, given in \cite{Bertacca:2014dra} (see also \cite{Bertacca:2014wga,Yoo:2014sfa,DiDio:2014lka,Bertacca:2014hwa}). Our results hold for arbitrary triangle shapes in Fourier space, and generalize the partial result of \cite{Kehagias:2015tda}, which applied a separate universe analysis to compute the relativistic effects in the special case of a squeezed bispectrum.

\subsection*{  Galaxy bispectrum in general relativity}

The observed galaxy number density contrast at redshift $z$ and in direction $\mathbf{n}$ is denoted by $\Delta_g(z,\mathbf{n})$. Since it is observable, $\Delta_g$ is a gauge-independent quantity, and any gauge can be used to compute it.
We only consider correlations at the same redshift. At fixed redshift $z$, the galaxy 3-point correlation function depends on $\bm n_i$ ($i=1,2,3$) and can be computed as the bispectrum in Fourier space at fixed {conformal time} $\eta(z)$, with {$\bm x =(\eta_0-\eta(z))\bm n$.} The galaxy bispectrum $B_g$ is defined by (suppressing the redshift dependence)
\begin{equation} \label{b1}
\langle \Delta_{g}( \mathbf{k}_{1}) \Delta_{g}( \mathbf{k}_{2}) \Delta_{g}( \mathbf{k}_{3}) \rangle = (2\pi)^{3}B_{g}(  \mathbf{k}_{1},  \mathbf{k}_{2},  \mathbf{k}_{3})\delta^{D}( \mathbf{k}_{1}+ \mathbf{k}_{2}+ \mathbf{k}_{3}).
\end{equation}

We assume Gaussian initial conditions and use a local univariate model of galaxy bias \cite{Baldauf:2010vn}:
\be\label{b12}
\delta_g(\bm x)=b_1 \delta_m(\bm x)+{1\over2}b_2\big[\delta_m(\bm x)^2-\big\langle
\delta_m(\bm x)^2\big\rangle \big].
\ee
We use Poisson gauge, but $\delta_g, \delta_m$ are the number and density contrasts in {total matter gauge},  in order to define bias correctly \cite{Jolicoeur:2016,Jeong:2011as}. {Here we assume scale-independent bias, but we allow for redshift dependence.}
We require $\Delta_g=\Delta_{g}^{(1)}+\Delta_{g}^{(2)}/2$ at second order if we want to compute $B_g$ to leading order:
\begin{align} \label{b2}
2\langle \Delta_{g}(\mathbf{k}_{1}) \Delta_{g}( \mathbf{k}_{2}) \Delta_{g}( \mathbf{k}_{3}) \rangle &=  \langle \Delta_{g}^{(1)}( \mathbf{k}_{1}) \Delta_{g}^{(1)}( \mathbf{k}_{2}) \Delta_{g}^{(2)}(\mathbf{k}_{3}) \rangle + \text{2 cyclic permutations}\,.
\end{align}
The expression that we use for $\Delta_g(z,\mathbf{n})$, including all general relativistic (GR) effects up to second order, is given in  \cite{Bertacca:2014dra}.
We neglect the integrated terms, and also the effects on  $\Delta_g$ of source evolution bias and magnification bias.

At first order
\begin{align} \label{b3}
\Delta_{g}^{(1)}(\mathbf{k}) &= \mathcal{K}^{(1)}( {k}, \mu)\delta_{m}^{(1)}(\mathbf{k}),
\end{align}
where the observed direction $\bm n$ appears through $\mu=\hat{\bm k}\cdot\bm n$. The kernel $\mathcal{K}^{(1)}({k}, \mu)$  maps the dark matter density contrast to the observed galaxy number density contrast.  
We split it into a Newtonian part $\mathcal{K}^{(1)}_{\rm{N}}$ and a GR correction $\mathcal{K}^{(1)}_{\rm{GR}}$. In the Newtonian part we include the redshift-space distortions:
\be
\mathcal{K}^{(1)}_{\rm{N}}({k},\mu) = b_{1} + f\mu^{2} \,,
\ee
where $f=-{\ud}\ln D/{\ud}\ln (1+z)$ is the linear growth rate. The correction from {GR local lightcone effects} is \cite{Challinor:2011bk,Jeong:2011as}
\bea
\mathcal{K}^{(1)}_{\rm{GR}}({k},\mu) &=&{ - {\mathrm{i} f}
\left[\frac{2}{ {\chi\H}}+\frac{\mathcal{H}'}{\mathcal{H}^2}  \right]\mu{\H \over k}+
{\left[3f+ {3\over2}\Omega_{m} \left(2-f - \frac{2}{\chi\H}-\frac{\mathcal{H}'}{\H^{2}} \right)\right]} \frac{\mathcal{H}^{2}}{k^{{2}}}\equiv {\rm i}\gamma_1{\mu\over k}+ {\gamma_2\over k^2},} \label{grk1} 
\eea
{where $\chi=\eta_0-\eta$.} The $\H/k$  term arises from Doppler effects, and the $(\H/k)^{2}$ term is due to gravitational potentials {(we have used the GR Poisson and Euler equations)}. 

At second order 
\begin{align}\label{dg2}
\Delta_{g}^{(2)}( \mathbf{k}) &= \int \frac{\ud^{3}k_{1}}{(2\pi)^{3}}\int \ud^{3}k_{2}\,
{\mathcal{K}^{(2)}( \mathbf{k}_{1}, \mathbf{k}_{2}, \k)}\delta_{m}^{(1)}( \mathbf{k}_{1})\delta_{m}^{(1)}( \mathbf{k}_{2})\delta^{D}(\mathbf{k}_{1} + \mathbf{k}_{2} - \mathbf{k}).
\end{align}
We again split the kernel into Newtonian and GR correction terms, each of which consists of evolution terms and projection effects. We neglect the GR corrections to the evolution of the density contrast since our focus is on projection effects (see \cite{Biern:2014zja,Villa:2015ppa,Hwang:2015jja} for the GR dynamical corrections). The Newtonian part is 
\be \label{kn2}
{\mathcal{K}_{\rm{N}}^{(2)}(\mathbf{k}_{1},  \mathbf{k}_{2},\k_3) = b_{1}F_{2}(\mathbf{k}_{1}, \mathbf{k}_{2}) + b_{2} + f\,G_{2}(\mathbf{k}_{1}, \mathbf{k}_{2})\mu_3^{2}+ {\cal Z}_2({\k_{1}},  {\k_{2}}), }
\ee
where $\mu_i=\hat\k_i\cdot\bm n$. {The first 3 terms on the right are dynamical and the last term is projection. The kernels in \eqref{kn2} are given by} \cite{Scoccimarro:1997st, Verde:1998zr} 
 \begin{eqnarray}
F_2({\k}_1,{\k}_2)&=&\frac{10}{7}+\frac{{\k}_1 \cdot {\k}_2}{k_1
k_2}\left(\frac{k_1}{k_2}+\frac{k_2}{k_1}\right)+\frac{4}{7}
\left(\frac{{\k}_1 \cdot {\k}_2}{k_1 k_2}\right)^2, \label{f2}\\
G_2({\k}_1,{\k}_2)&=& \frac{6}{7} + \frac{{\k}_1\cdot {\k}_2}{k_1k_2}\left(\frac{k_1}{k_2}+\frac{k_2}{k_1}\right) +\frac{8}{7} \left(\frac{{\k}_1\cdot {\k}_2}{k_1 k_2}\right)^2, \label{g2}\\
{{\cal Z}_2({\k_{1}},  {\k_{2}})}&=& {f^2{\mu_1\mu_2 \over k_1k_2}\big( \mu_1k_1+\mu_2k_2\big)^2+
b_1{f\over k_1k_2}\left[ \big(\mu_1^2+\mu_2^2 \big)k_1k_2+\mu_1\mu_2\big(k_1^2+k_2^2 \big) \right].}
%2f^{2}\mu_{{1}}^{2}\mu_{{2}}^{2}  + b_{1}f\left(\mu_{{1}}^{2} + \mu_{{2}}^{2}\right)
%+ f^{2}\frac{\big(\mu_{{1}}^{3}\mu_{{2}} k_{1}^{2}+ \mu_{{2}}^{3}\mu_{{1}} k_{2}^{2}\big)}{k_{1}k_{2}}  + b_{1}f\mu_{{1}}\mu_{{2}}\frac{\big(k_{1}^{2} + k_{2}^{2}\big)}{k_{1}k_{2}}.
\end{eqnarray}
These kernels describe respectively the nonlinear effects from the density contrast, peculiar velocities and redshift-space distortions. {Equations \eqref{f2} and \eqref{g2} apply to a matter-dominated cosmology, but the corrections induced by a cosmological constant are small \cite{Bernardeau:2001qr}.} 
Note that the overall scaling with $k$ for $\mathcal{K}_{\rm{N}}^{(2)}$ is $\mathcal{O}(k^0)$: only the angles between the observed direction $\bm n$ and $\k_i$  appear.
\begin{figure}[htbp]
\begin{center}
\includegraphics[width=0.49\textwidth]{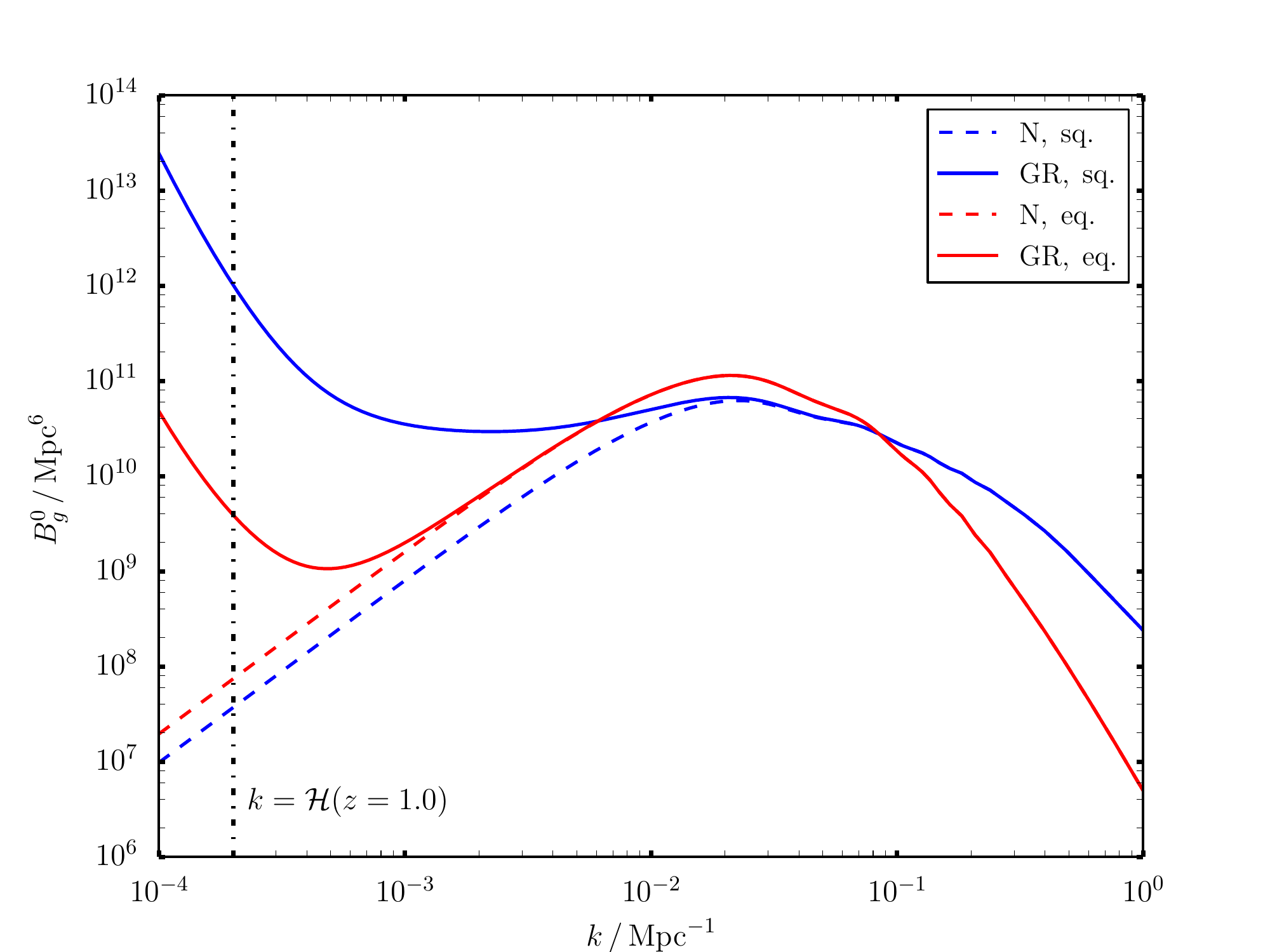}
\includegraphics[width=0.49\textwidth]{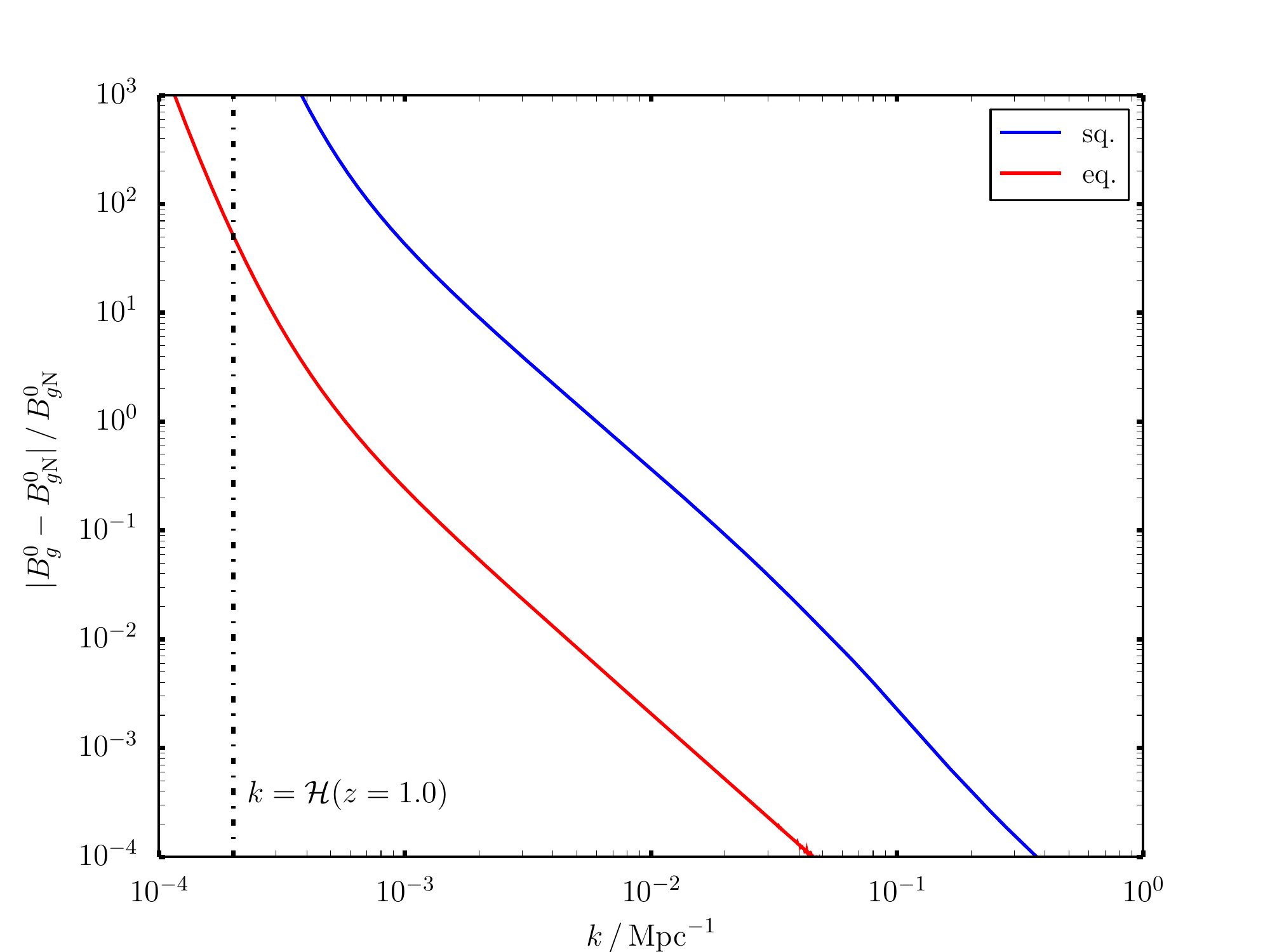}
\caption{ {\em Left:} Monopole of the bispectrum including all {GR local lightcone} corrections (GR, solid) compared to its Newtonian approximation (N, dashed), for equilateral (eq., red) and squeezed (sq., blue) triangles. {\em Right:} Fractional difference due to GR corrections. (Note: the Hubble constant is fixed at $h=0.678$.)  }
\label{kjascdjksncsajkncda}
\end{center}
\end{figure}

The computation of the kernel for  {GR local lightcone effects} is a key result of this work (see \cite{Jolicoeur:2016} for further details). {We start with $\Delta_g^{(2)}$, as given in Eq. (19) of \cite{Bertacca:2014dra}. We neglect the integrated terms and set the source evolution bias $b_e$ to zero, and then transform this expression into Fourier space. We also transform $\delta_g^{(2)}$ from Poisson to total matter gauge, in order to apply the bias relation \eqref{b12}. Then we express each term as proportional to $\delta_{m}^{(1)}( \mathbf{k}_{1})\delta_{m}^{(1)}( \mathbf{k}_{2})$, using the Einstein and conservation equations. The result is:}
%\footnote{
%{$\Gamma_2$ in arXiv v2 has been absorbed into $\Gamma_1$; the previous $\Gamma_I, I=3,\cdots,14$ have been re-labelled $I=2,\cdots,13$, and the new $\Gamma_{14}$ term here is a correction to arXiv v2.}}   
{\begin{align}\label{kjdsncskcnsjkdn}
\mathcal{K}^{(2)}_{\mathrm{GR}}(\bm{k}_{1}, \bm{k}_{2}, \bm{k}_{3}) &= \frac{1}{k_{1}^{2}k_{2}^{2}}\bigg\{\Gamma_{1} + {\rm i}\left(\mu_{1}k_{1} + \mu_{2}k_{2}\right)\Gamma_{2} + \bigg(\frac{k_{1}k_{2}}{k_{3}}\bigg)^{2}\bigg[F_{2}(\bm{k}_{1}, \bm{k}_{2})\Gamma_{3} + G_{2}(\bm{k}_{1}, \bm{k}_{2})\Gamma_{4} \bigg] \nonumber \\
 &~~~+ \mu_{1}\mu_{2}k_{1}k_{2}\,\Gamma_{5} + \left(\bm{k}_{1}\cdot \bm{k}_{2}\right)\Gamma_{6} + \left(k_{1}^{2} + k_{2}^{2}\right)\Gamma_{7} + (\mu_{1}^{2}k_{1}^{2} + \mu_{2}^{2}k_{2}^{2})\Gamma_{8} 
\nonumber \\
&~~~ +  {\rm i}\bigg[\left(\mu_{1}k_{1}^{3} + \mu_{2}k_{2}^{3}\right)\Gamma_{9}  + \left(\mu_{1}k_{1} + \mu_{2}k_{2}\right)\left(\bm{k}_{1} \cdot \bm{k}_{2}\right)\Gamma_{10}  + k_{1}k_{2}\left(\mu_{1}k_{2} + \mu_{2}k_{1}\right)\Gamma_{11}  
\nonumber \\
&~~~+ {(\mu_{1}^{3}k_{1}^{3}+\mu_{2}^{3}k_{2}^{3})}\,\Gamma_{12}  + \mu_{1}\mu_{2}k_{1}k_{2}\left(\mu_{1}k_{1} + \mu_{2}k_{2}\right)\Gamma_{13} + \mu_{3}\frac{k_{1}^{2}k_{2}^{2}}{k_{3}}G_{2}(\bm{k}_{1}, \bm{k}_{2})\Gamma_{14}\bigg] \bigg\},
    \end{align}}
where $\Gamma_I(z)$ are given in the Appendix.
In contrast to the Newtonian kernel, the GR correction terms  have  scalings  from $\mathcal{O}(k^{-4})$ to $\mathcal{O}(k^{-1})$,  so that they are suppressed on small scales.  
The leading order GR {local lightcone} corrections \eqref{grk1} to the power spectrum are linear and are confined to ultra-large scales. By contrast, the leading order GR {local lightcone} corrections to the bispectrum are nonlinear -- and mode coupling between linear GR terms and Newtonian terms means that the GR corrections in the bispectrum are present on smaller scales than in the power spectrum.

Using Wick's theorem, the bispectrum is given in terms of the kernels by
\begin{align} \label{b6}
B_{g}( \mathbf{k}_{1},  \mathbf{k}_{2},  \k_3) &= { \mathcal{K}^{(1)}( \k_{1})\mathcal{K}^{(1)}({\k}_{2}) \mathcal{K}^{(2)}(  \mathbf{k}_{1},  \mathbf{k}_{2}, \k_3)}
P(k_{1})P(k_{2}) +\text{2 cyclic permutations}\,,
\end{align}
where $P(k)$ is the linear matter power spectrum.

In the squeezed limit, we have {
\be
k_1\approx k_2=k_S\gg k_L=k_3,\qquad \mu_2\approx -\mu_1,~~ \mu_3 \approx \sqrt{1-\mu_1^2}\,\cos\phi, 
\ee
where $\phi$ is the azimuthal angle. 
We average \eqref{b6} over $\phi$, to get $\big\langle B_{g}^{\rm sq}\big\rangle_\phi=(1/2\pi)\int_0^{2\pi}{\rm d}\phi\,B_{g}^{\rm sq}$. Then we average $\big\langle B_{g}^{\rm sq}\big\rangle_\phi$ over $\mu_1$ to give the monopole, $B_g^{{\rm sq}\,0}=(1/2)\int_{-1}^{+1}{\rm d}\mu_1 \,\big\langle B_{g}^{\rm sq}\big\rangle_\phi$. This leads to (see \cite{Maartens:2016} for further details):
\bea
B_{g}^{{\rm sq}\,0}  &=& \bigg({{\cal B}_2\over k_L^2} + {{\cal B}_4\over k_L^4} \bigg)P(k_S)P(k_L),
\label{bgsm} \\
{\cal B}_2 &=& 
 {2\over105}\Bigg\{{\gamma_2}\Big(80b_1f+21b_1f^2+
 150b_1^2 +35b_1^2f  +105b_1b_2+ 35b_2f+18f^2\Big)
+7b_1\Big[5\big(3b_1+f \big)
\Gamma_7+ \big(5b_1+3f \big)\Gamma_{8}\Big] \nonumber\\
&+& {\gamma_2}\Big(35b_1^2f+21b_1f^2+6f^3\Big)-\gamma_1\Big[\big(35b_1+7f\big)\Gamma_{11}+\big(7b_1+3f\big)\Gamma_{13}\Big]+f\Big[\big(35b_1+7f\big)\Gamma_{7}+\big(7b_1+3f\big)\Gamma_{8}\Big]
\nonumber\\
    &{-}& {\gamma_1}\Big[\big(7b_1+3f \big)f\gamma_1+7\big(5b_1+f\big)\Gamma_9+3\big(7b_1+f\big)\Gamma_{12}\Big]   \Bigg\}, 
\label{ca2} \\
{\cal B}_4 &=& {2\over15} \gamma_2  \Big[5\big(3b_1+f\big)\Gamma_7+\big(5b_1+3f\big)\Gamma_{8}
-{\gamma_1}\big(5\Gamma_9 + 3 \Gamma_{12}\big)
\Big], 
  \label{ca4} 
\eea 
where $\gamma_2$ is defined by \eqref{grk1}. }

In order to illustrate the nature and magnitude of the {GR local lightcone corrections} to the bispectrum, we choose an isosceles configuration, with $k_1=k_2\equiv k$ and $k_3= k\sqrt{2(1-\cos\theta_{12})}$. We consider the cases of equilateral ($\theta_{12}=\pi/3$) and squeezed ($\theta_{12}\ll1$) triangles, and we set the  redshift to $z=1$, {for which the comoving distance from the observer to the galaxies is $\chi_g\approx 3.4\,$Gpc.}  For the galaxy bias, {we follow \cite{Pollack:2013alj}} and take $b_1=\sqrt{1+z}$ and $b_2=-0.3\sqrt{1+z}$. {The cosmological parameters are those from Planck 2015 \cite{Ade:2015xua}.}
In Fig.~\ref{kjascdjksncsajkncda}, the monopole of the bispectrum with {GR local lightcone corrections} is compared to the Newtonian approximation.  {Note that the results on ultra-large transverse scales, $k_\perp\lesssim \chi_g^{-1}/10 \sim (340\,{\rm Mpc})^{-1}$, will be modified by wide-angle correlations, which have been neglected in the plane-parallel approximation.}

\vspace*{0.2cm}

\underline{\em Squeezed shapes:} We choose  $\cos\theta_{12}=0.998$ so that $k_3\approx k/16$. Even for this moderately squeezed case we see a significant departure from the Newtonian prediction when the short-wavelength sides $k$ (long sides in Fourier space) approach the equality scale: the difference is  $\gtrsim 10\%$ for $k\lesssim (50\,{\rm Mpc})^{-1}$. {Because of mode coupling, the GR correction reaches percent-level at  surprisingly small scales, below the equality scale.  
If we take the current Hubble scale as an observable upper limit for the triangle sides,  the long-wavelength side is a maximum  when $k_3=H_0\approx 2.3\times10^{-4}\,$Mpc$^{-1}$, corresponding to {$k\approx 16k_3\approx (270\,{\rm Mpc})^{-1}$.} At this maximum, the GR {bispectrum} is {$\sim {3.4}$ times} the Newtonian {prediction}.

\vspace*{0.2cm}

\underline{\em Equilateral shapes:} We see a similar behaviour, {with a smaller difference of $\sim{0.1}\%$ at equality scales} and {$\gtrsim {23}\%$} when $k=k_3\lesssim  1\,$Gpc$^{-1}$. The GR correction reaches percent-level at {$k\sim ({225}\,{\rm Mpc})^{-1}$}.
The equal wavelengths of the triangle are at the observable maximum when $k=k_3=H_0 $. At this maximum, the GR {bispectrum} is {$\sim {33}$} times the Newtonian {prediction}.

\vspace*{0.2cm}

{We define the reduced monopole of the bispectrum (including {GR local lightcone corrections}) as}
\begin{equation}\label{eq:reducedbispectrum}
Q({\k}_1,{\k}_2,{\k}_3) = \frac{{B_g^0}({\k}_1,{\k}_2,{\k}_3)}{\big[{P_g^0(\k_1}) {P_g^0(\k_2}) +  \text{2 cyclic permutations}\big]}\,,
\end{equation}
where {$P_{g}^0$} is the {monopole of the} linear galaxy power spectrum including GR corrections at first order. In the Newtonian approximation, {$B_g^0$} in \eqref{eq:reducedbispectrum} is replaced by {$B_{g,{\rm N}}^0$} and {$P_g^0$} is replaced by the {monopole of the} Newtonian galaxy power spectrum (including redshift-space distortions).

The dimensionless  {reduced monopole of the} bispectrum, as a function of the angle $\theta_{12}$, is shown in
 Fig.~\ref{kjascdjksncsajkcsjkcnsjkdcsncda}, for scales  $k\approx k_{\rm eq}$ and $k= 1\,$Gpc$^{-1}$. 
In the equality-scale case,  the fractional GR correction is $>1\%$ for {$\theta_{12} /\pi\lesssim0.1$}.
For the gigaparsec scale,
the fractional GR correction is $\gtrsim{10}\%$ for all $\theta_{12}$, reaching a maximum of $\sim {60}\%$. 
\begin{figure}[htbp]
\begin{center}
\includegraphics[width=0.49\textwidth]{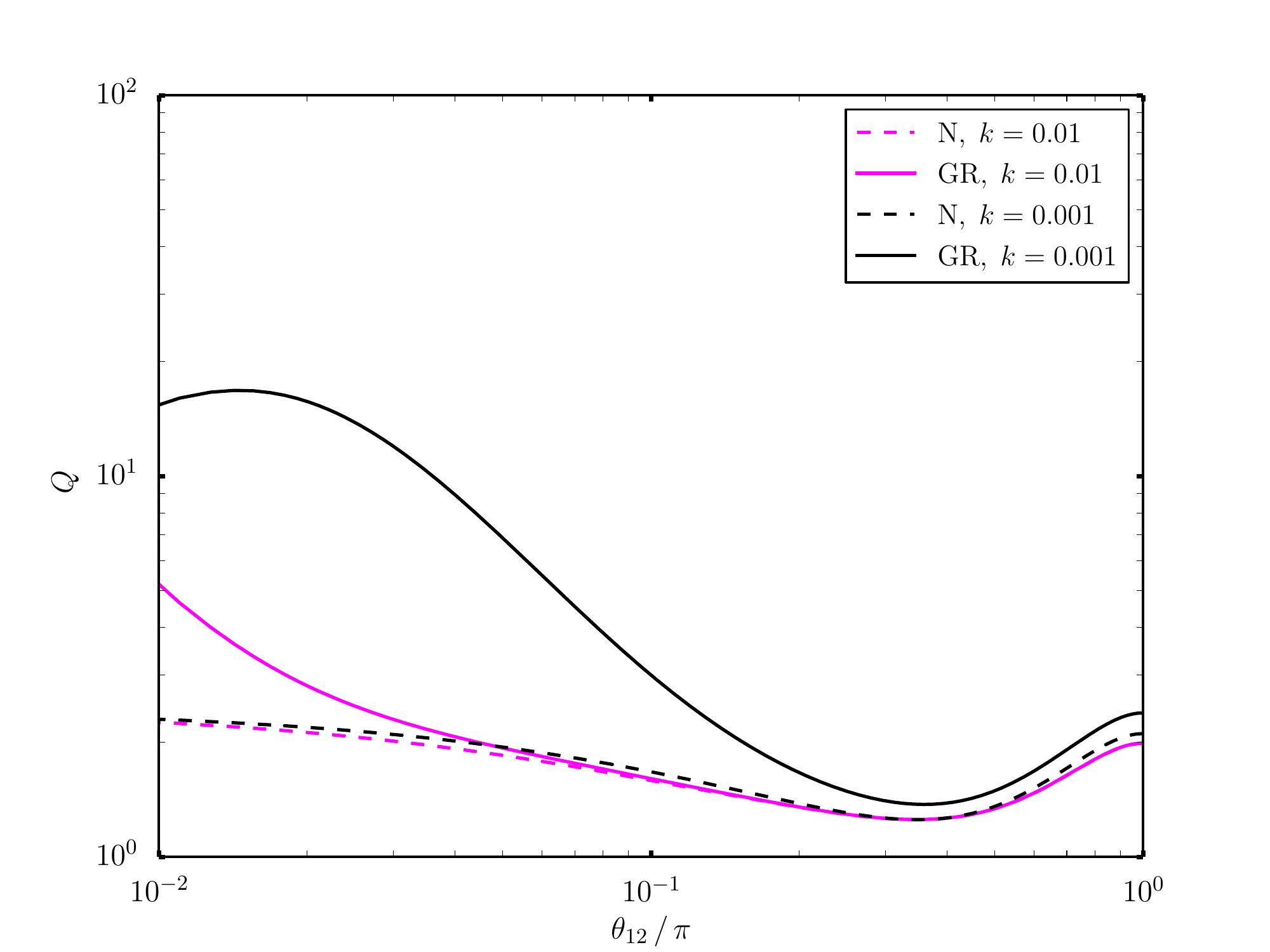}
\includegraphics[width=0.49\textwidth]{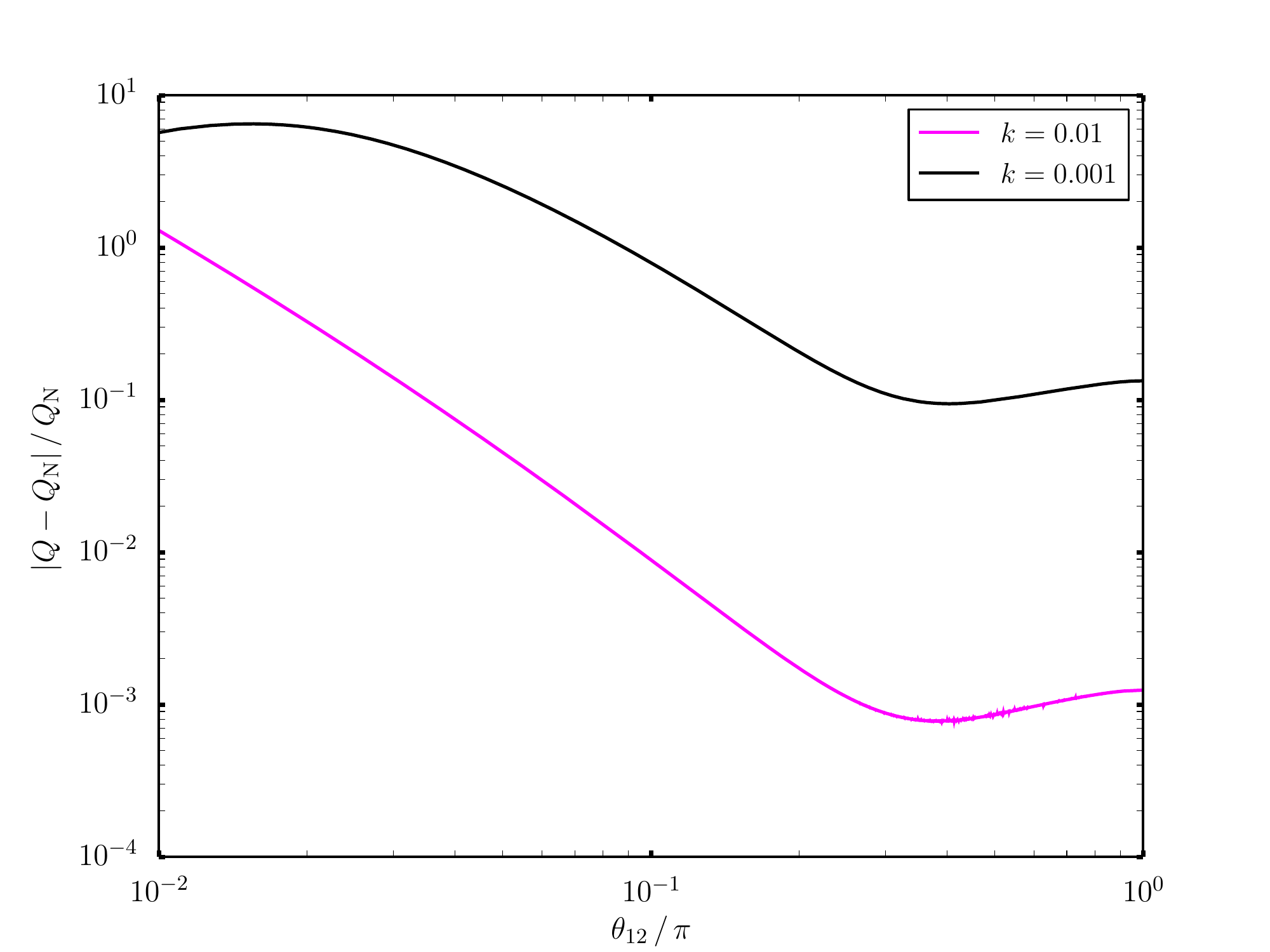}
\caption{{\em Left:}  Reduced monopole of the bispectrum {with GR local lightcone corrections} (GR, solid) compared to the Newtonian approximation (N, dashed),  as a function of the isosceles angle,  for equality-scale (magenta) and Gpc-scale (black) wavelengths of the equal sides $k$.  {\em Right:} Fractional correction from GR effects. }
\label{kjascdjksncsajkcsjkcnsjkdcsncda}
\end{center}
\end{figure}

\newpage
{\subsection*{ Misinterpreting the primordial universe} }

We see from Figs.~\ref{kjascdjksncsajkncda} and \ref{kjascdjksncsajkcsjkcnsjkdcsncda} how {GR local projection effects}, in a Gaussian primordial universe, boost the large and ultra-large scale power in the bispectrum. This behaviour is qualitatively similar to the effect of primordial non-Gaussianity on the Newtonian bispectrum. In a Newtonian analysis, one needs to subtract off the nonlinear effects from evolution and from redshift-space distortions in order to isolate the primordial non-Gaussian signal. Our results show that {GR local projection effects} contribute  a significant contamination -- a Newtonian analysis of the bispectrum, in the case where equality scales and larger are being probed, could be seriously misleading by ignoring these GR projection effects.  \\

For a simple illustration, we consider the effect of local primordial non-Gaussianity on the Newtonian galaxy bispectrum, but we neglect the scale-dependence in the galaxy bias. The primordial gravitational potential is \begin{equation}\label{eq:localform}
\Phi^\text{nG}(\bm x)=\varphi(\bm x) + \fnl  \left(\varphi(\bm x)^2 - \big\langle\varphi(\bm x)^2\big\rangle\right) ,
\end{equation}
where $\varphi$ is a first-order Gaussian potential. As a consequence, the matter density contrast receives a correction:
\begin{equation}\label{eq:gaussianize}
\delta_m^\text{nG}(z,{\k})=
\alpha(z,k)\left[ \varphi({\k})+ \fnl \int \frac{{\rm d}^3k'}{(2\pi)^3}
\varphi({\bm k'})\varphi({\k}-{\k'})\right]
~~~\text{with}~~~
\alpha(k,z)=\frac{2 k^2  D(z) T(k)}{  3H_0^2
 \Omega_{m0}}  
\frac{g(z=0)}{g(z_\infty)},
\end{equation}
where $T$ is the transfer function ($T\to 1$ for $k\to 0$), and $g=(1+z)D$ is the growth suppression factor.
Then a Newtonian analysis of the bispectrum, neglecting scale-dependence in the galaxy bias, leads to
\bea 
B_{g,{\rm N}}^{\rm nG}( \mathbf{k}_{1},  \mathbf{k}_{2},  \k_{3}) &=& 
{\mathcal{K}^{(1)}_{\rm{N}}( \mathbf{k}_{1})\mathcal{K}^{(1)}_{\rm{N}}( \mathbf{k}_{2})\Bigg[ \mathcal{K}^{(2)}_{\rm{N}}( \mathbf{k}_{1},  \mathbf{k}_{2}, \k_{3}) } { + 2\fnl\mathcal{K}^{(1)}_{\rm{N}}(\mathbf{k}_{3}) \frac{\alpha(k_3)}{\alpha(k_1)\alpha(k_2)}\Bigg] P(k_{1})P(k_{2})}  \nonumber\\ 
&&{}+  \text{2 cyclic permutations}\,.
\label{eq:NewtonianbispecttrumnG}
\eea

Suppose that  we interpret the observed galaxy bispectrum  using the standard Newtonian analysis.  In order to fit the observations we would match the theoretical bispectrum~\eqref{eq:NewtonianbispecttrumnG} to the observed bispectrum. However, the observed bispectrum {necessarily} includes the GR corrections from projection effects. Therefore, we would effectively be matching~\eqref{eq:NewtonianbispecttrumnG} to  \eqref{b6}. 
This would {produce} an effective $\fnl^\text{eff}$:
\bea
\fnl^\text{eff}(\k_1,\k_2,\k_3)&=& \bigg\{\left[\mathcal{K}^{(1)}_{\rm{GR}}(\mathbf{k}_{1})\mathcal{K}^{(1)}(\mathbf{k}_{2}) {\mathcal{K}^{(2)}( \mathbf{k}_{1},  \mathbf{k}_{2},\k_3) }
+\mathcal{K}^{(1)}_{\rm{N}}(\mathbf{k}_{1})\mathcal{K}^{(1)}_{\rm{GR}}(\mathbf{k}_{2}){\mathcal{K}^{(2)}( \mathbf{k}_{1},  \mathbf{k}_{2},\k_3) }\right.\nonumber\\&& \left.
{}+\mathcal{K}^{(1)}_{\rm{N}}(\mathbf{k}_{1})\mathcal{K}^{(1)}_{\rm{N}}(\mathbf{k}_{2}){\mathcal{K}^{(2)}_{\rm GR}( \mathbf{k}_{1},  \mathbf{k}_{2},\k_3) }\right]P(k_{1})P(k_{2}) +  \text{2 cyclic permutations}\bigg\}
\nonumber\\&&
\times \bigg\{2\mathcal{K}^{(1)}_{\rm{N}}(\mathbf{k}_{1}) \mathcal{K}^{(1)}_{\rm{N}}(\mathbf{k}_{2})\mathcal{K}^{(1)}_{\rm{N}}(\mathbf{k}_{3}) \frac{\alpha(k_3)}{\alpha(k_1)\alpha(k_2)} P(k_{1})P(k_{2}) +  \text{2 cyclic permutations}
\bigg\}^{-1}. 
\eea

\begin{figure}[htbp]
\begin{center}
\includegraphics[width=0.49\textwidth]{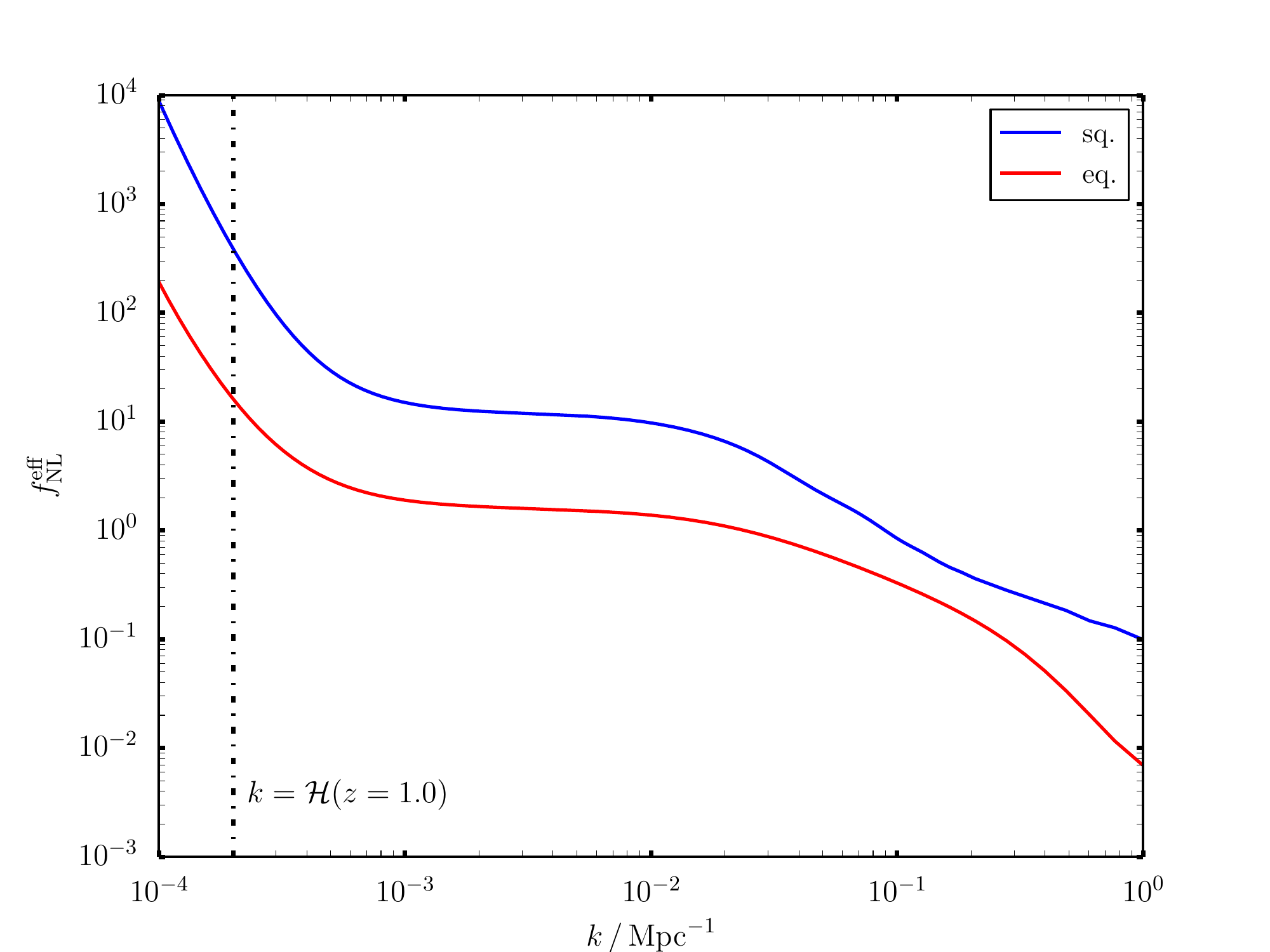}
\caption{ Effective non-Gaussianity due to {GR local projection effects}, from comparing to a Newtonian analysis, but neglecting scale-dependent galaxy bias. (Curves correspond to the cases considered in Fig.~\ref{kjascdjksncsajkncda}.) }
\label{jfdadhjfsbvdfsbhvdsjhbvds}
\end{center}
\end{figure}

Figure~\ref{jfdadhjfsbvdfsbhvdsjhbvds} shows $\fnl^\text{eff}$. At equality scales, $\fnl^\text{eff}\sim {1.4}$ for the equilateral and $\sim {9.0}$ for the squeezed case.
On the false basis of a Newtonian interpretation, we would conclude that the primordial universe was significantly non-Gaussian and that the non-Gaussianity was of nonlocal type. 
Of course, we have neglected the scale-dependent galaxy bias that is induced by
local primordial non-Gaussianity, {which will reduce $\fnl^\text{eff}$.} This scale-dependent bias produces similar effects to the GR lightcone effects  -- which reinforces our conclusion that the GR lightcone effects need to be included for a consistent analysis.\\

\subsection*{ Conclusions}

Non-Gaussianity in the galaxy distribution arises from three effects -- primordial, dynamical and projection effects.   
The standard Newtonian analysis of the bispectrum incorporates the projection effect due to redshift-space distortions,  {but omits all other projection effects, as well as the relativistic corrections to nonlinear dynamics.}  This has been extended by \cite{DiDio:2015bua} to include the GR projection effect of weak lensing, {but they neglected all other GR projection and dynamical effects}. {Our work complements theirs by computing for the first time the galaxy bispectrum in a primordial Gaussian universe, with all local GR projection effects included, up to second order. We also neglect the GR dynamical corrections, and we follow the standard approach of working in Fourier space, that uses the plane-parallel approximation, which affects our results on ultra-large transverse scales, $k_\perp\lesssim  (340\,{\rm Mpc})^{-1}$.}

Our key results are the nonlinear GR correction kernel \eqref{kjdsncskcnsjkdn} and the analytical form of the squeezed limit of the galaxy bispectrum \eqref{bgsm}. {GR local lightcone effects} in the galaxy bispectrum for isosceles shapes are shown in Figs.~\ref{kjascdjksncsajkncda} and \ref{kjascdjksncsajkcsjkcnsjkdcsncda}.

A major aim of future galaxy surveys is to measure primordial non-Gaussianity with precision down to $|\fnl|<1$, beyond the capabilities of CMB experiments, using both the power spectrum \cite{Alonso:2015sfa,Fonseca:2015laa} and bispectrum \cite{Tellarini:2016sgp}. 
{It is known that neglecting GR projection effects in the galaxy power spectrum can bias the measurement of $\fnl$ \cite{Bruni:2011ta,Camera:2014sba}.
We have extended this to the galaxy bispectrum, showing how linear and nonlinear {GR local projection effects} will contaminate the primordial non-Gaussian signal. The mode coupling present at leading order in the bispectrum means that the GR effects can be non-negligible at smaller scales than for the power spectrum. 

The strongest effects arise for squeezed shapes: with moderate squeezing, the GR correction to the bispectrum is {$\sim {10}\%$ of the Newtonian prediction when the short-wavelength sides are $k\sim (50\,{\rm Mpc})^{-1}$}.  For equilateral shapes, the GR correction to the bispectrum reaches percent-level at {$k\sim ({225}\,{\rm Mpc})^{-1}$}.  

{We have taken the first step towards a complete analysis of ultra-large scale GR effects on the galaxy bispectrum. Our results are incomplete, and further work is needed to incorporate the omitted GR effects -- nonlocal projection,  wide-angle and dynamical. However our partial results already show that it is essential for an accurate measurement of $\fnl$ to incorporate the GR effects in theoretical analysis of the galaxy bispectrum for next-generation surveys.}

\[\]
{\bf\small Acknowledgments:}\\
We thank Daniele Bertacca, Ruth Durrer, Kazuya Koyama, Sabino Matarrese and David Wands for useful discussions and comments.
All authors are funded in part by the NRF (South Africa). OU, SJ and RM are also supported by the South African SKA Project. RM is also supported by the UK STFC, Grants  ST/K00090X/1 and ST/N000668/1.

%\appendix

\newpage
\subsection*{Appendix: Redshift evolution of the GR correction kernel }

The {$\Gamma_I(z)$} functions appearing in the GR correction kernel~\eqref{kjdsncskcnsjkdn} are given by \cite{Jolicoeur:2016}:
%\footnote{{$\Gamma_2$ in arXiv v2 has been absorbed into $\Gamma_1$; the previous $\Gamma_I, I=3,\cdots,14$ have been re-labelled $I=2,\cdots,13$,  and the new $\Gamma_{14}$ term here is a correction to arXiv v2.}}  
\begin{align}\label{A1}
\frac{\Gamma_{1}}{\mathcal{H}^{4}} &= \frac{9}{4}\Omega_{m}^{2}\Bigg[ 6f-5 -2{f'\over\H} +{2\over\chi\H}\left( 4f-4+{1\over\chi\H}+3{\H'\over \H^2 }\right) +4(f-2){\H'\over \H^2 } +3{\H^{\prime 2}\over \H^4}- {\H''\over \H^3 } \Bigg]
\nonumber \\
&~~~ +\frac{9}{2}\Omega_{m}f\left(11-2f-\frac{4}{\chi\H}-\frac{\H'}{\H^{2}}\right) +3f^{2}\left(4-\frac{\H'}{\H^{2}}\right) \nonumber \\
%------------------------------
\frac{\Gamma_{2}}{\H^{3}} &=   \frac{3}{2}\Omega_{m}f \Bigg[(f-1)(2-f) -{f'\over\H} +{2\over\chi\H}\left( 2f+{1\over\chi\H}+3{\H'\over \H^2 }\right) +(2f-3){\H'\over \H^2 } +3{\H^{\prime 2}\over \H^4}- {\H''\over \H^3 } \Bigg]
\nonumber \\
&~~~ -\frac{9}{4}\Omega_{m}^{2} -3f\bigg(1+\frac{2}{\chi\H}\bigg) \nonumber \\
%------------------------------
\frac{\Gamma_{3}}{\H^{2}} &= \frac{3}{2}\Omega_{m}\left(1-2f+\frac{2}{\chi\H}-3\frac{\H'}{\H^{2}}\right) - \frac{3\Omega_{m}'}{2\H} \nonumber \\
%------------------------------
\frac{\Gamma_{4}}{\H^{2}} &= 3f \nonumber \\
%------------------------------
%\frac{\Gamma_{5}}{\H^{2}} &= 3\Omega_{m}f(2-f) + f^{2} \Bigg[ 4-{2\over\chi\H}\left( 4+{1\over\chi\H}+3{\H'\over \H^2 }\right) -3{\H'\over \H^2 } -3{\H^{\prime 2}\over \H^4}+ {\H''\over \H^3 } \Bigg]
%\nonumber \\
%------------------------------
%\frac{\Gamma_{6}}{\H^{2}} &= f^{2}\left(1+\frac{\H'}{\H^{2}}\right) + 3\Omega_{m}f \nonumber \\
\frac{\Gamma_{5}}{\H^{2}} &= 3\Omega_{m}f(2-f) + f^{2} \Bigg[ 4-{2\over\chi\H}\left( {3}+{1\over\chi\H}+3{\H'\over \H^2 }\right) -3{\H'\over \H^2 } -3{\H^{\prime 2}\over \H^4}+ {\H''\over \H^3 } \Bigg]
\nonumber \\
\frac{\Gamma_{6}}{\H^{2}} &= f^{2}\left(1 {-\frac{2}{\chi\H}}+ \frac{\H'}{\H^{2}}\right) + 3\Omega_{m}f
\nonumber \\
\frac{\Gamma_{7}}{\H^{2}} &= \frac{3}{2}\Omega_{m}\Bigg[b_{1}\bigg(2-\frac{2}{\chi\H}-\frac{\H'}{\H^{2}}\bigg) + \frac{b_{1}'}{\H} \Bigg] + f^{2}\Bigg[b_{1}(3-f)-\frac{b_{1}'}{\H}\Bigg] \nonumber \\
%------------------------------
\frac{\Gamma_{8}}{\H^{2}} &= \frac{9}{4}\Omega_{m}^{2} + \frac{3}{2}\Omega_{m}f\left(3-2f-\frac{4}{\chi\H}-\frac{\H'}{\H^{2}}\right) + 3f^{2} \nonumber \\
%------------------------------
\frac{\Gamma_{9}}{\H} &= -\frac{3}{2}\Omega_{m}b_{1} \nonumber \\
%------------------------------
\frac{\Gamma_{10}}{\H} &= f^{2} \nonumber \\
%------------------------------
\frac{\Gamma_{11}}{\H} &= f\Bigg[b_{1}\bigg(f-\frac{2}{\chi\H}-\frac{\H'}{\H^{2}}\bigg) + \frac{b_{1}'}{\H} \Bigg] - f^{2} \nonumber \\
%------------------------------
\frac{\Gamma_{12}}{\H} &=- \frac{3}{2}\Omega_{m}f \nonumber \\
%------------------------------
\frac{\Gamma_{13}}{\H} &=\frac{3}{2}\Omega_{m}f - f^{2}\left(3+\frac{4}{\chi\H}+\frac{3\H'}{\H^{2}}\right)  \nonumber \\
%------------------------------
\frac{\Gamma_{14}}{\H} &= -f\left(\frac{2}{\chi\H}+\frac{\H'}{\H^{2}}\right) \nonumber
\end{align}


\begin{thebibliography}{99}

%\cite{Tellarini:2016sgp}
\bibitem{Tellarini:2016sgp} 
  M.~Tellarini, A.~J.~Ross, G.~Tasinato and D.~Wands,
  %``Galaxy bispectrum, primordial non-Gaussianity and redshift space distortions,''
  JCAP {\bf 1606}, 014 (2016)
  %doi:10.1088/1475-7516/2016/06/014
  [arXiv:1603.06814].
  %%CITATION = doi:10.1088/1475-7516/2016/06/014;%%
  %3 citations counted in INSPIRE as of 11 Oct 2016

%\cite{DiDio:2015bua}
\bibitem{DiDio:2015bua} 
  E.~Di Dio, R.~Durrer, G.~Marozzi and F.~Montanari,
  %``The bispectrum of relativistic galaxy number counts,''
  JCAP {\bf 1601}, 016 (2016)
  %doi:10.1088/1475-7516/2016/01/016
  [arXiv:1510.04202].
  %%CITATION = doi:10.1088/1475-7516/2016/01/016;%%
  %8 citations counted in INSPIRE as of 11 Oct 2016

%\cite{Verde:1998zr}
\bibitem{Verde:1998zr} 
  L.~Verde, A.~F.~Heavens, S.~Matarrese and L.~Moscardini,
  %``Large scale bias in the universe. 2. Redshift space bispectrum,''
  Mon.\ Not.\ Roy.\ Astron.\ Soc.\  {\bf 300}, 747 (1998)
 % doi:10.1046/j.1365-8711.1998.01937.x
  [astro-ph/9806028].
  %%CITATION = doi:10.1046/j.1365-8711.1998.01937.x;%%
  %52 citations counted in INSPIRE as of 11 Oct 2016

%\cite{Baldauf:2010vn}
\bibitem{Baldauf:2010vn} 
  T.~Baldauf, U.~Seljak and L.~Senatore,
  %``Primordial non-Gaussianity in the Bispectrum of the Halo Density Field,''
  JCAP {\bf 1104}, 006 (2011)
  %doi:10.1088/1475-7516/2011/04/006
  [arXiv:1011.1513].
  %%CITATION = doi:10.1088/1475-7516/2011/04/006;%%
  %45 citations counted in INSPIRE as of 11 Oct 2016

{
%\cite{Gil-Marin:2016wya}
\bibitem{Gil-Marin:2016wya} 
  H.~Gil-Mar\'in, W.~J.~Percival, L.~Verde, J.~R.~Brownstein, C.~H.~Chuang, F.~S.~Kitaura, S.~A.~Rodr\'iguez-Torres and M.~D.~Olmstead,
  %``The clustering of galaxies in the SDSS-III Baryon Oscillation Spectroscopic Survey: RSD measurement from the power spectrum and bispectrum of the DR12 BOSS galaxies,''
Mon.\ Not.\ Roy.\ Astron.\ Soc.\  {\bf 465}, 1757 (2016) 
  [arXiv:1606.00439].
}

\bibitem{Jolicoeur:2016}
	 S.~Jolicoeur, O.~Umeh, R.~Maartens and C.~Clarkson,
  %``in preparation,''
   in preparation (2016).
  %%CITATION = doi:10.1088/1475-7516/2014/09/037;%%
  %26 citations counted in INSPIRE as of 11 Oct 2016

%\cite{Bertacca:2014dra}
\bibitem{Bertacca:2014dra} 
  D.~Bertacca, R.~Maartens and C.~Clarkson,
  %``Observed galaxy number counts on the lightcone up to second order: I. Main result,''
  JCAP {\bf 1409}, 037 (2014)
 % doi:10.1088/1475-7516/2014/09/037
  [arXiv:1405.4403].
  %%CITATION = doi:10.1088/1475-7516/2014/09/037;%%
  %26 citations counted in INSPIRE as of 11 Oct 2016


%\cite{Bertacca:2014wga}
\bibitem{Bertacca:2014wga} 
  D.~Bertacca, R.~Maartens and C.~Clarkson,
  %``Observed galaxy number counts on the lightcone up to second order: II. Derivation,''
  JCAP {\bf 1411},  013 (2014)
 % doi:10.1088/1475-7516/2014/11/013
  [arXiv:1406.0319].
  %%CITATION = doi:10.1088/1475-7516/2014/11/013;%%
  %24 citations counted in INSPIRE as of 11 Oct 2016


%\cite{Yoo:2014sfa}
\bibitem{Yoo:2014sfa} 
  J.~Yoo and M.~Zaldarriaga,
  %``Beyond the Linear-Order Relativistic Effect in Galaxy Clustering: Second-Order Gauge-Invariant Formalism,''
  Phys.\ Rev.\ D {\bf 90},  023513 (2014)
  %doi:10.1103/PhysRevD.90.023513
  [arXiv:1406.4140].
  %%CITATION = doi:10.1103/PhysRevD.90.023513;%%
  %27 citations counted in INSPIRE as of 11 Oct 2016


%\cite{DiDio:2014lka}
\bibitem{DiDio:2014lka} 
  E.~Di Dio, R.~Durrer, G.~Marozzi and F.~Montanari,
  %``Galaxy number counts to second order and their bispectrum,''
  JCAP {\bf 1412}, 017 (2014)
  [Erratum: JCAP {\bf 1506},  E01 (2015)]
 % doi:10.1088/1475-7516/2014/12/017, 10.1088/1475-7516/2015/06/E01
  [arXiv:1407.0376].
  %%CITATION = doi:10.1088/1475-7516/2014/12/017, 10.1088/1475-7516/2015/06/E01;%%
  %32 citations counted in INSPIRE as of 11 Oct 2016


%\cite{Bertacca:2014hwa}
\bibitem{Bertacca:2014hwa} 
  D.~Bertacca,
  %``Observed galaxy number counts on the light cone up to second order: III. Magnification bias,''
  Class.\ Quant.\ Grav.\  {\bf 32}, 195011 (2015)
 % doi:10.1088/0264-9381/32/19/195011
  [arXiv:1409.2024].
  %%CITATION = doi:10.1088/0264-9381/32/19/195011;%%
  %10 citations counted in INSPIRE as of 11 Oct 2016


%\cite{Kehagias:2015tda}
\bibitem{Kehagias:2015tda} 
  A.~Kehagias, A.~M.~Dizgah, J.~Nore\~na, H.~Perrier and A.~Riotto,
  %``A Consistency Relation for the Observed Galaxy Bispectrum and the Local non-Gaussianity from Relativistic Corrections,''
  JCAP {\bf 1508},  018 (2015)
 % doi:10.1088/1475-7516/2015/08/018
  [arXiv:1503.04467].
  %%CITATION = doi:10.1088/1475-7516/2015/08/018;%%
  %8 citations counted in INSPIRE as of 11 Oct 2016


%\cite{Jeong:2011as}
\bibitem{Jeong:2011as} 
  D.~Jeong, F.~Schmidt and C.~M.~Hirata,
  %``Large-scale clustering of galaxies in general relativity,''
  Phys.\ Rev.\ D {\bf 85}, 023504 (2012)
  %doi:10.1103/PhysRevD.85.023504
  [arXiv:1107.5427].
  %%CITATION = doi:10.1103/PhysRevD.85.023504;%%
  %94 citations counted in INSPIRE as of 11 Oct 2016


%\cite{Challinor:2011bk}
\bibitem{Challinor:2011bk} 
  A.~Challinor and A.~Lewis,
  %``The linear power spectrum of observed source number counts,''
  Phys.\ Rev.\ D {\bf 84}, 043516 (2011)
 % doi:10.1103/PhysRevD.84.043516
  [arXiv:1105.5292].
  %%CITATION = doi:10.1103/PhysRevD.84.043516;%%
  %166 citations counted in INSPIRE as of 11 Oct 2016


%\cite{Biern:2014zja}
\bibitem{Biern:2014zja} 
  S.~G.~Biern, J.~O.~Gong and D.~Jeong,
  %``Nonlinear matter bispectrum in general relativity,''
  Phys.\ Rev.\ D {\bf 89}, 103523 (2014)
 % doi:10.1103/PhysRevD.89.103523
  [arXiv:1403.0438].
  %%CITATION = doi:10.1103/PhysRevD.89.103523;%%
  %4 citations counted in INSPIRE as of 11 Oct 2016


%\cite{Villa:2015ppa}
\bibitem{Villa:2015ppa} 
  E.~Villa and C.~Rampf,
  %``Relativistic perturbations in $\Lambda$CDM: Eulerian & Lagrangian approaches,''
  JCAP {\bf 1601},  030 (2016)
  %doi:10.1088/1475-7516/2016/01/030
  [arXiv:1505.04782 [gr-qc]].
  %%CITATION = doi:10.1088/1475-7516/2016/01/030;%%
  %6 citations counted in INSPIRE as of 11 Oct 2016


%\cite{Hwang:2015jja}
\bibitem{Hwang:2015jja} 
  J.~c.~Hwang, D.~Jeong and H.~Noh,
  %``Cosmological non-linear density and velocity power spectra including non-linear vector and tensor modes,''
  Mon.\ Not.\ Roy.\ Astron.\ Soc.\  {\bf 459},  1124 (2016)
  %doi:10.1093/mnras/stw621
  [arXiv:1509.07534].
  %%CITATION = doi:10.1093/mnras/stw621;%%
  %2 citations counted in INSPIRE as of 11 Oct 2016


%\cite{Scoccimarro:1997st}
\bibitem{Scoccimarro:1997st} 
  R.~Scoccimarro, S.~Colombi, J.~N.~Fry, J.~A.~Frieman, E.~Hivon and A.~Melott,
  %``Nonlinear evolution of the bispectrum of cosmological perturbations,''
  Astrophys.\ J.\  {\bf 496}, 586 (1998)
 % doi:10.1086/305399
  [astro-ph/9704075].
  %%CITATION = doi:10.1086/305399;%%
  %153 citations counted in INSPIRE as of 11 Oct 2016

{
%\cite{Bernardeau:2001qr}
\bibitem{Bernardeau:2001qr} 
  F.~Bernardeau, S.~Colombi, E.~Gaztanaga and R.~Scoccimarro,
  %``Large scale structure of the universe and cosmological perturbation theory,''
  Phys.\ Rept.\  {\bf 367}, 1 (2002)
  [astro-ph/0112551].
}

\bibitem{Maartens:2016}
R. Maartens, K. Koyama, D. Wands, O. Umeh, C. Clarkson and S. Jolicoeur, in preparation (2016).

{
%\cite{Pollack:2013alj}
\bibitem{Pollack:2013alj} 
  J.~E.~Pollack, R.~E.~Smith and C.~Porciani,
  %``A new method to measure galaxy bias,''
  Mon.\ Not.\ Roy.\ Astron.\ Soc.\  {\bf 440},  555 (2014)
  [arXiv:1309.0504].
}

%\cite{Ade:2015xua}
\bibitem{Ade:2015xua} 
  P.~A.~R.~Ade {\it et al.} [Planck Collaboration],
  %``Planck 2015 results. XIII. Cosmological parameters,''
  Astron.\ Astrophys.\  {\bf 594}, A13 (2016)
  %doi:10.1051/0004-6361/201525830
  [arXiv:1502.01589].
  %%CITATION = doi:10.1051/0004-6361/201525830;%%
  %2245 citations counted in INSPIRE as of 11 Oct 2016


%\cite{Alonso:2015sfa}
\bibitem{Alonso:2015sfa} 
  D.~Alonso and P.~G.~Ferreira,
  %``Constraining ultralarge-scale cosmology with multiple tracers in optical and radio surveys,''
  Phys.\ Rev.\ D {\bf 92},  063525 (2015)
 % doi:10.1103/PhysRevD.92.063525
  [arXiv:1507.03550].
  %%CITATION = doi:10.1103/PhysRevD.92.063525;%%
  %20 citations counted in INSPIRE as of 11 Oct 2016


%\cite{Fonseca:2015laa}
\bibitem{Fonseca:2015laa} 
  J.~Fonseca, S.~Camera, M.~Santos and R.~Maartens,
  %``Hunting down horizon-scale effects with multi-wavelength surveys,''
  Astrophys.\ J.\  {\bf 812},  L22 (2015)
  %doi:10.1088/2041-8205/812/2/L22
  [arXiv:1507.04605].
  %%CITATION = doi:10.1088/2041-8205/812/2/L22;%%
  %12 citations counted in INSPIRE as of 11 Oct 2016


%\cite{Bruni:2011ta}
\bibitem{Bruni:2011ta} 
  M.~Bruni, R.~Crittenden, K.~Koyama, R.~Maartens, C.~Pitrou and D.~Wands,
  %``Disentangling non-Gaussianity, bias and GR effects in the galaxy distribution,''
  Phys.\ Rev.\ D {\bf 85}, 041301 (2012)
  %doi:10.1103/PhysRevD.85.041301
  [arXiv:1106.3999].
  %%CITATION = doi:10.1103/PhysRevD.85.041301;%%
  %58 citations counted in INSPIRE as of 11 Oct 2016


%\cite{Camera:2014sba}
\bibitem{Camera:2014sba} 
  S.~Camera, R.~Maartens and M.~G.~Santos,
  %``Einstein's legacy in galaxy surveys,''
  Mon.\ Not.\ Roy.\ Astron.\ Soc.\  {\bf 451},  L80 (2015)
 % doi:10.1093/mnrasl/slv069
  [arXiv:1412.4781].
  %%CITATION = doi:10.1093/mnrasl/slv069;%%
  %17 citations counted in INSPIRE as of 11 Oct 2016
  
\end{thebibliography}
\end{document}